\documentclass[epj,a4paper,nofootinbib,nobibnotes,twocolumn,groupedaddress,floatfix]{svjour}
\usepackage{graphicx}
\usepackage[utf8]{inputenc}  
\usepackage[T1]{fontenc}
\usepackage{color}
\usepackage{amsmath}
\usepackage{amssymb}
\usepackage{dcolumn}
\usepackage[numbers,sort&compress]{natbib}
\usepackage{setspace}
\usepackage{tabularx}
\usepackage[caption=false]{subfig} 
\usepackage{booktabs,array}
\usepackage[colorlinks,linkcolor=blue,citecolor=blue,urlcolor=blue,bookmarks=false,hypertexnames=true]{hyperref}
\usepackage[flushleft]{threeparttable}
\usepackage{verbatim,multirow}
\usepackage{siunitx}
\usepackage{tablefootnote}

\begin{document}

\title{First direct limit on the 334 keV resonance strength in $^{22}$Ne($\alpha$,$\gamma$)$^{26}$Mg reaction}

\author{D.\,Piatti \inst{1,2,3} 
\and  E.\,Masha \inst{4,5}
%
\and  M.\,Aliotta\inst{6}
%
\and J.\,Balibrea-Correa\inst{7}
%
\and F. Barile\inst{14,15}
%
\and  D.\,Bemmerer\inst{3} 
\thanks{d.bemmerer@hzdr.de}
\and  A.\,Best\inst{7,8}
%
\and  A.\,Boeltzig\inst{9,10}
%
\and  C.\,Broggini\inst{1}
%
\and  C.G.\,Bruno\inst{6}
%
\and  A.\,Caciolli\inst{1,2}
\thanks{caciolli@pd.infn.it}
\and  F.\,Cavanna\inst{11}
%
\and T.\,Chillery\inst{6}
%
\and G.\,F.\,Ciani\inst{9,10}
%
\and A.\,Compagnucci\inst{1,2}
%
\and P.\,Corvisiero\inst{12,13}
%
\and L.\,Csedreki\inst{9,10,16}
%
\and T.\,Davinson\inst{6}
%
\and R.\,Depalo\inst{1,2}
%
\and A.\,di\,Leva\inst{7,8}
%
\and Z.\,Elekes\inst{16}
%
\and F.\,Ferraro\inst{12,13}
%
\and E.\,M.\,Fiore\inst{14,15}
\and A.\,Formicola\inst{9,10}
%
\and Zs.\,Fülöp\inst{16}
%
\and G.\,Gervino\inst{11,17}
%
\and A.\,Guglielmetti\inst{4,5}
%
\and C.\,Gustavino\inst{18}
%
\and Gy.\,Gyürky\inst{16}
%
\and G.\,Imbriani\inst{7,8}
%
\and M.\,Junker\inst{9,10}
%
\and M.\,Lugaro\inst{19,20}
%
\and P.\,Marigo\inst{1,2}
%
\and R.\,Menegazzo\inst{1}
%
\and V.\,Mossa\inst{14,15}
%
\and F. R.\,Pantaleo\inst{14,15}
%
\and V.\,Paticchio\inst{15}
%
\and R.\,Perrino\inst{15}
\thanks{Permanent address: INFN Sezione di Lecce, Lecce, Italy}
\and P.\,Prati\inst{12,13}
%
\and D.\,Rapagnani\inst{7}
%
\and L.\,Schiavulli\inst{14,15}
%
\and J.\,Skowronski\inst{1,2}
%
\and K.\,Stöckel\inst{3,21}
%
\and O.\,Straniero\inst{9,22}
%
\and T.\,Szücs\inst{3}
%
\and M.\,P.\,Takács\inst{3,21}\thanks{Present address: Physikalisch-Technische Bundesanstalt, Bundesallee 100, 38116 Braunschweig, Germany}
%
\and  S.\,Zavatarelli\inst{13}
%
}

\institute{INFN, Sezione di Padova, Via F. Marzolo 8, 35131 Padova, Italy
\and
Università degli Studi di Padova, Via F. Marzolo 8, 35131 Padova, Italy
\and
Helmholtz-Zentrum Dresden-Rossendorf, Bautzner Landstraße 400, 01328 Dresden, Germany
\and
Università degli Studi di Milano, Via G. Celoria 16, 20133 Milano, Italy
\and
INFN, Sezione di Milano, Via G. Celoria 16, 20133 Milano, Italy
\and
SUPA, Scottish Universities Physics Alliance, School of Physics and Astronomy, University of Edinburgh, EH9 3FD Edinburgh, United Kingdom
\and
Università degli Studi di Napoli "Federico II", Dipartimento di Fisica "E. Pancini", Via Cintia, 80126 Napoli, Italy
\and
INFN, Sezione di Napoli, Via Cintia, 80126 Napoli, Italy
\and
Gran Sasso Science Institute, 67100 L'Aquila, Italy
\and
INFN Laboratori Nazionali del Gran Sasso (LNGS), 67100 Assergi (AQ), Italy
\and
INFN, Sezione di Torino, Via P. Giuria 1, 10125 Torino, Italy
\and
Università degli Studi di Genova, Via Dodecaneso 33, 16146 Genova, Italy
\and
Istituto Nazionale di Fisica Nucleare (INFN), Sezione di Genova, Via Dodecaneso 33, 16146 Genova, Italy
\and
Università degli Studi di Bari, 70125 Bari, Italy
\and
INFN, Sezione di Bari, 70125 Bari, Italy
\and
Institute for Nuclear Research (ATOMKI), PO Box 51, H-4001 Debrecen, Hungary
\and
Università degli Studi di Torino, Via P. Giuria 1, 10125 Torino, Italy
\and
INFN, Sezione di Roma, Piazzale A. Moro 2, 00185 Roma, Italy
\and
Konkoly Observatory, Research Centre for Astronomy and Earth Sciences, Hungarian Academy of Sciences, 1121 Budapest, Hungary
\and
ELTE E\"otv\"os Lor\'and University, Institute of Physics, Budapest 1117, P\'azm\'any P\'eter s\'et\'any 1/A, Hungary
\and
Technische Universität Dresden, Institut für Kern- und Teilchenphysik, Zellescher Weg 19, 01069 Dresden, Germany
\and
INAF-Osservatorio Astronomico d'Abruzzo, Via Mentore Maggini , 64100, Teramo, Italy
}

\date{\today} 

\abstract{
In stars, the fusion of $^{22}$Ne and $^4$He may produce either $^{25}$Mg, with the emission of a neutron, or $^{26}$Mg and a $\gamma$ ray. 
At high temperature, the ($\alpha,n$) channel dominates, while at low temperature, it is energetically hampered. The rate of its competitor, the $^{22}$Ne($\alpha$,$\gamma$)$^{26}$Mg reaction, and, hence, the minimum temperature for the ($\alpha,n$) dominance, are controlled by many nuclear resonances. The strengths of these resonances have hitherto been studied only indirectly.
The present work aims to directly measure the total strength of the resonance at $E$\textsubscript{r}$\,=\,$334$\,$keV (corresponding to $E$\textsubscript{x}$\,=\,$10949$\,$keV in $^{26}$Mg). 
The data reported here have been obtained using high intensity $^4$He$^+$ beam from the INFN LUNA 400 kV underground accelerator, a windowless, recirculating, 99.9\% isotopically enriched $^{22}$Ne gas target, and a 4$\pi$ bismuth germanate summing $\gamma$-ray detector. The ultra-low background rate of less than 0.5 counts/day was determined using 67 days of no-beam data and 7 days of $^4$He$^+$ beam on an inert argon target.
The new high-sensitivity setup allowed to determine the first direct upper limit of 4.0$\,\times\,$10$^{-11}$ eV (at 90\% confidence level) for the resonance strength. 
Finally, the sensitivity of this setup paves the way to study further $^{22}$Ne($\alpha$,$\gamma$)$^{26}$Mg resonances at higher energy. 
}%

\titlerunning{Experimental limit on the 10949 keV resonance strength in $^{22}$Ne($\alpha$,$\gamma$)$^{26}$Mg}

\authorrunning{Denise Piatti et al. (LUNA)}

\maketitle

\section{Introduction}

The $^{22}$Ne + $\alpha$ fusion reactions $^{22}$Ne($\alpha$,$\gamma$)$^{26}$Mg and \linebreak $^{22}$Ne($\alpha$,$n$)$^{25}$Mg  impact several astrophysical scenarios \cite{Longland12-PRC, The-2007Apj, Karakas06-ApJ}. Both the ($\alpha,n$) and the ($\alpha,\gamma$) reaction rates are dominated by a number of resonances, most of which have hitherto only been investigated via indirect methods. In the present work, the high sensitivity setup and the analysis method leading to the first experimental upper limit are reported for the resonance at 334 keV center-of-mass energy in the $^{22}$Ne($\alpha,\gamma$)$^{26}$Mg reaction. This introduction reviews first the astrophysical and then the nuclear aspects of the $^{22}$Ne($\alpha,\gamma$)$^{26}$Mg  reaction. 

\subsection{Astrophysical aspects of $^{22}$Ne($\alpha,\gamma$)$^{26}$Mg}

The stable neon isotope $^{22}$Ne (with 9.25\% isotopic abundance in the Solar System) is synthesized in stars via the following $\alpha$-chain:
\begin{equation}
 ^{14}{\rm N}(\alpha,\gamma)^{18}{\rm F}(\beta ^{+}\,\nu)^{18}{\rm O}(\alpha,\gamma)^{22}{\rm Ne} , 
 \label{eq:14n}
\end{equation} 
which operates in He-burning regions (both in central cores and in shells) 
at typical temperatures above $T \approx 100$ MK.
The destruction of $^{22}$Ne, instead, occurs mainly via two competing channels \cite{Clayton03-Book} with different $Q$ values \cite{Wang-2017ChPhC}:
\begin{eqnarray}
& ^{22}{\rm Ne}(\alpha,\gamma)^{26}{\rm Mg} & Q\,=\,10614.74(3)\,{\rm keV} , \label{eq:ne22ag} \\
& ^{22}{\rm Ne}(\alpha,n)^{25}{\rm Mg} & Q\,=\,-478.34(5)\,{\rm keV} . \label{eq:ne22an}
\end{eqnarray}
At high temperatures ($T \gtrsim 300$ MK), the ($\alpha,n$) reaction (\ref{eq:ne22an}) dominates.  At lower temperatures ($T < 300$ MK) instead, this channel is energetically inhibited because of its negative $Q$ value, and only the ($\alpha,\gamma$) reaction (\ref{eq:ne22ag}) remains active. The exact cross-over temperature at which the ($\alpha,n$) rate exceeds the ($\alpha,\gamma$) one depends on the details of the reaction cross sections (\ref{eq:ne22ag}, \ref{eq:ne22an}) and represents a crucial parameter for a number of stellar scenario models \cite{Longland12-PRC, The-2007Apj, Karakas06-ApJ}.

In massive stars \cite[$M_{\rm i} \gtrsim 10\, M_{\odot}$;][]{Raiteri_etal_91, Pignatari10-ApJ} the $^{22}$Ne($\alpha, n$)$^{25}$Mg is the predominant neutron source for the \textit{slow} neutron capture process, the so-called astrophysical \textit{s}-process. It occurs in convective He-burning cores and in the subsequent convective C-burning shells. On the other hand the release of neutrons through the ($\alpha,n$) reaction plays a minor role for the \textit{s}-process in thermally-pulsing asymptotic giant branch (TP-AGB) stars of intermediate mass  \cite[$M_{\rm i} \gtrsim 3-8\, M_{\odot}$;][]{Wiescher12-ARAA, vanRaai12-AA, Karakas_etal_06, Busso_etal_99}, when they undergo He-shell flashes. 
Moreover, significant amounts of the neutron-rich isotopes $^{25,26}$Mg are produced by the $^{22}$Ne+$\alpha$ reactions during the TP-AGB phase of intermediate mass stars \cite{Ventura_etal_18, Cristallo-2015ApJ, Karakas-2014PASA, Marigo_etal13, Herwig05-ARAA}. 
The competition between the two channels and, hence, the magnesium isotopic ratio depend critically on temperature at the base of the convective region induced by the thermal pulse (pulse-driven convective zone, hereinafter also PDCZ). 

In addition the abundances of the magnesium isotopes at the stellar surface may be significantly affected by both the third dredge-up and the Hot-Bottom Burning nuclesynthesis \cite[e.g.,][]{ForestiniCharbonnel_97}.

The nuclear reactions (\ref{eq:ne22ag}) and (\ref{eq:ne22an}), active in the PDCZ, can impact the chemical enrichment of the two magnesium isotopes $^{25, 26}$Mg, both in the form of gas ejecta \cite{Slemer17-MNRAS, Cristallo-2015ApJ, Karakas06-ApJ}  or trapped into silicate-type dust grains \cite{Nanni_etal_14, FerrarottiGail_06}. 
Therefore the magnesium isotopic abundances in the stellar ejecta is relevant for various astrophysical questions, from the interpretation of the meteoritic pre-solar dust grains \cite{Zinner_2005GeCoA} and spectral analyses of cool stars in the Galaxy \cite{Melendez_2007ApJ, Yong03-ApJ}, to  the possibility of placing constraints and characterizing the first-generation stellar polluters responsible for the chemical Mg-Al anticorrelation and Mg isotopic ratios measured in stars of globular clusters \cite{Ventura_etal_18,Melendez-2009ApJ, DaCosta-2009ApJ, Yong-2006ApJ, Yong03-AA}, and, in general, of constraining the chemical enrichment of galaxies \cite{Melendez-2009ApJ,Fenner03-PASA,Yong_etal_03}.

\subsection{Nuclear physics of $^{22}$Ne($\alpha,\gamma$)$^{26}$Mg}

The nuclear aspects of the $^{22}$Ne + $\alpha$ reactions have been studied by a number of different methods \cite[and references therein]{Adsley-21, Longland12-PRC}. The center-of-mass energies $E$ for the scenarios discussed above are in the $250 \le E \le 900$ keV range, roughly corresponding to $^{26}$Mg excitation energies $10865 \le E_{\rm x} \le 11515$ keV. There, for the $^{22}$Ne($\alpha,\gamma$)$^{26}$Mg, and in a smaller range due to the neutron threshold $S_n$ = 11093.08(4) keV \cite{Wang-2017ChPhC} for the $^{22}$Ne($\alpha,n$)$^{25}$Mg reaction both reaction rates are dominated by a number of nuclear resonances. Most of them are known by spectroscopic study results. The only exception is the $E_{\rm x}$ = 11319(2) keV ($E$ = 704(2) keV) resonance \cite{Jaeger01-PRL}, which has been studied directly both in the ($\alpha,n$) \cite{Jaeger01-PRL} and ($\alpha,\gamma$) \cite{Wolke89-ZPA, Hunt-2019PhRvC} channels, but with only limited precision.
The precise excitation energies and spin-parity assignments of many additional excited states in the $^{26}$Mg compound nucleus in the relevant energy range are still under debate. These additional levels that may strongly contribute to the $^{22}$Ne($\alpha,\gamma$)$^{26}$Mg reaction rate include the $E$ = 334.4(8), 469(1), and 556.33(5) keV resonances corresponding to excited levels $E_{\rm x}$ (spin parity $J^\pi$) = 10949.1(1) keV (1$^-$), 11084(1) keV (2$^+$), and 11171.07(4)keV (2$^+$), respectively \footnote{We adopted the $E_{\rm x}$ and $J^\pi$ values suggested in a recent review \cite{Adsley-21}.}.
The last resonance lies above the neutron threshold and thus affects both $^{22}$Ne+$\alpha$ reactions. The lower-lying resonances affect only the $^{22}$Ne($\alpha,\gamma$)$^{26}$Mg reaction rate, but they are quite uncertain, as illustrated by data from two recent independent studies using the $^{22}$Ne($^6{\rm Li},d$)$^{26}$Mg reaction. For the 469 keV resonance strength, \cite{Talwar16-PRC} report an upper limit of $\le 2.95 \times 10^{-11}$ eV, while \cite{Jayatissa-2020PhLB} provide a value of $(2.8 \pm 0.8) \times 10^{-10}$ eV. For the 556 keV strength instead, \cite{Talwar16-PRC} report a value of $(5.4 \pm 0.7) \times 10^{-7}$ eV and a natural spin parity of 1$^-$, while \cite{Jayatissa-2020PhLB} give an upper limit of $\le 6.5 \times 10^{-11}$ eV and $J^\pi$ = 2$^+$. 
Considering also that for a different system in the same mass range, i.e. the $^{22}$Ne(p,$\gamma$)$^{23}$Na reaction, direct experiments provided interesting and relevant data \cite{Cavanna15-PRL,Bemmerer18-EPL,Ferraro18-PRLb}, re-investigating the $^{22}$Ne($\alpha,\gamma$)$^{26}$Mg case with the direct methods seems desirable.
\begin{table*}[bth]
\centering
\caption{\small{Literature data for the $E$ = 334\ keV resonance in the  $^{22}$Ne($\alpha$,$\gamma$)$^{26}$Mg reaction: excitation energy $E_{\rm x}$, spin-parity $J^\pi$, and resonance strength $\omega\gamma$. For those works that determined a value of $\omega\gamma$, the last column denotes the method adopted: Measurement of the $\alpha$-spectroscopic factor ($S_\alpha$) 
}\label{tab:soa}}
\begin{tabular}{p{32mm} l l l D{x}{\times}{-1} l}
\toprule
    Reference & $E$\textsubscript{x} [keV] & $E$ [keV] & $J^{\pi}$ & \multicolumn{1}{c}{$\omega\gamma$ [eV]} & $\omega\gamma$ method \\
\midrule
\cite{Giesen93-NPA} & 10949$\,\pm\,$25  & 338 & 3$^-$ & 1.7x 10^{-13} & $S_\alpha$ from $^{22}$Ne($^{6}$Li,d) \\
\cite{NACRE99-NPA} &  -  & 338.4$\,\pm\,1.7$ & 3$^-$ & \le 1.4x10^{-13} & Evaluation\\  
\cite{Ugalde07-PRC} &  10953$\,\pm\,$25 &  & 5$^-$, 6$^+$, 7$^-$ &\\  
\cite{Longland09-PRC} &  10949.1$\,\pm\,$0.8 &  & 1$^-$ & \\
\cite{Iliadis10-NPA841_251, Sallaska13-ApJSS} & -  & 334.31$\,\pm\,0.1$ & 1 &  \le 3.6x10^{-9} & Evaluation\tablefootnote{Here, the much lower and more recent value in Ref. \cite{Longland12-PRC} by many of the same authors is used, see subsequent line.} \\
\cite{Longland12-PRC} & 10949  & 334.30$\,\pm\,$0.15 & 1$^-$ & \le 8.7x10^{-15} & Evaluation \\
\cite{Talwar16-PRC} & 10951$\,\pm\,21$  & 336 & 1$^-$ & (2\pm1)x10^{-13} & $S_\alpha$ from $^{22}$Ne($^{6}$Li,d) \\
\cite{Lotay19-EPJA} & & 334 & 1$^{-}$ & 8.69x10^{-14} & $S_\alpha$ from previous works\\
\cite{Jayatissa-2020PhLB} & 10950$\,\pm\,$20 &  &  & (9.0\pm2.4)x10^{-14} & ANC from $^{22}$Ne($^{6}$Li,d) \\
\cite{Adsley-21} & 10949.1$\,\pm\,$0.1  & 334.4$\,\pm\,$0.8 & 1$^-$ & \le 8.7x10^{-15} & Evaluation\\
Present work &   
&  
&  
& \le 4.0x10^{-11} & Direct measurement\\
\bottomrule
\end{tabular}

\end{table*}

The aim of this work is to present a first step of just such a re-investigation, starting with the $E$ = 334 keV resonance that corresponds to the $^{26}$Mg excited level at $E_{\rm x}$ = 10949.1(8) keV \cite{Basunia-2016NDS}.  Due to its relatively isolated location in the level scheme, this resonance dominates the total reaction rate $N_{\rm A}\langle \sigma v \rangle$ at 100-200 MK temperature.

The contribution $N_{\rm A}\langle \sigma v \rangle_i$ of an isolated, narrow resonance $i$ with center of mass energy $E_i$ and strength $\omega\gamma_i$ to the thermonuclear reaction rate at temperature $T$ is directly proportional to $\omega\gamma_i$ and given by \cite{Iliadis15-Book}:
\begin{equation}\label{eq:TNRR}
N_{\rm A}\langle \sigma v \rangle_i = N_A \left( \frac{2\pi }{\mu k_{\rm B}T} \right)^\frac{3}{2} \hbar^2 \; \omega\gamma_i\; \exp \left[-\frac{E_i}{k_{\rm B}T} \right],
\end{equation}
with $\mu=m_1 m_2 / (m_1 + m_2)$ the reduced mass of nuclei $m_{1,2}$, $k_{\rm B}$ the Boltzmann constant and $\hbar$ reduced Planck's constant. The strength $\omega\gamma_i$, in turn, is defined as
\begin{eqnarray}
\omega\gamma_i & = & \frac{(2 J_i +1)}{(2J_\alpha + 1) (2J_{\rm Ne-22} + 1)} \frac{\Gamma_\alpha \Gamma_\gamma}{\Gamma_\alpha + \Gamma_\gamma} \\ 
    & = & (2 J_i +1) \frac{\Gamma_\alpha \Gamma_\gamma}{\Gamma_\alpha + \Gamma_\gamma},
\end{eqnarray}
with $J_{i}$, $J_{\alpha}$, and $J_{\rm Ne-22}$ the spins of resonance $i$, $\alpha$ beam and target nucleus, respectively. $\Gamma_{\alpha,\gamma}$ are the $\alpha$ and $\gamma$ widths of the corresponding excited state, respectively. 
For the case of the $E$ = 334 keV resonance, the excitation energy and $J^\pi$ = $1^-$ have been obtained using polarized photons \cite{Longland09-PRC}, and the $\gamma$ decay branchings of the level have been established with the ($\gamma,\gamma'$) method \cite{Longland09-PRC} and they are now adopted in \cite{Basunia-2016NDS}.
Its $\gamma$-width of $\Gamma_\gamma$ = 1.87(30) eV has been determined in a ($\gamma,\gamma'$) study \cite{deBoer10-PRC}, while $\Gamma_\alpha$ was found to be (3$\pm$1)$\times10^{-14}$ eV assuming a spin-parity of 1$^-$ in a recent ($^6$Li,d) spectroscopic factor measurement \cite{Jayatissa-2020PhLB}. The literature data are reviewed in Table ~\ref{tab:soa}.

This paper is organized as follows: the experimental setup and procedures are described in Section ~\ref{sec:Experiment}. Section ~\ref{sec:Analysis} deals with the data analysis and experimental results. In Section ~\ref{sec:Future}, the experimental sensitivity for other resonances is discussed. 
The summary and conclusion are given in Section ~\ref{sec:Summary}.


\section{Experiment} \label{sec:Experiment}

\begin{figure}
\centering
    \includegraphics[width=\columnwidth,trim=0cm 2cm 0cm 6cm,clip]{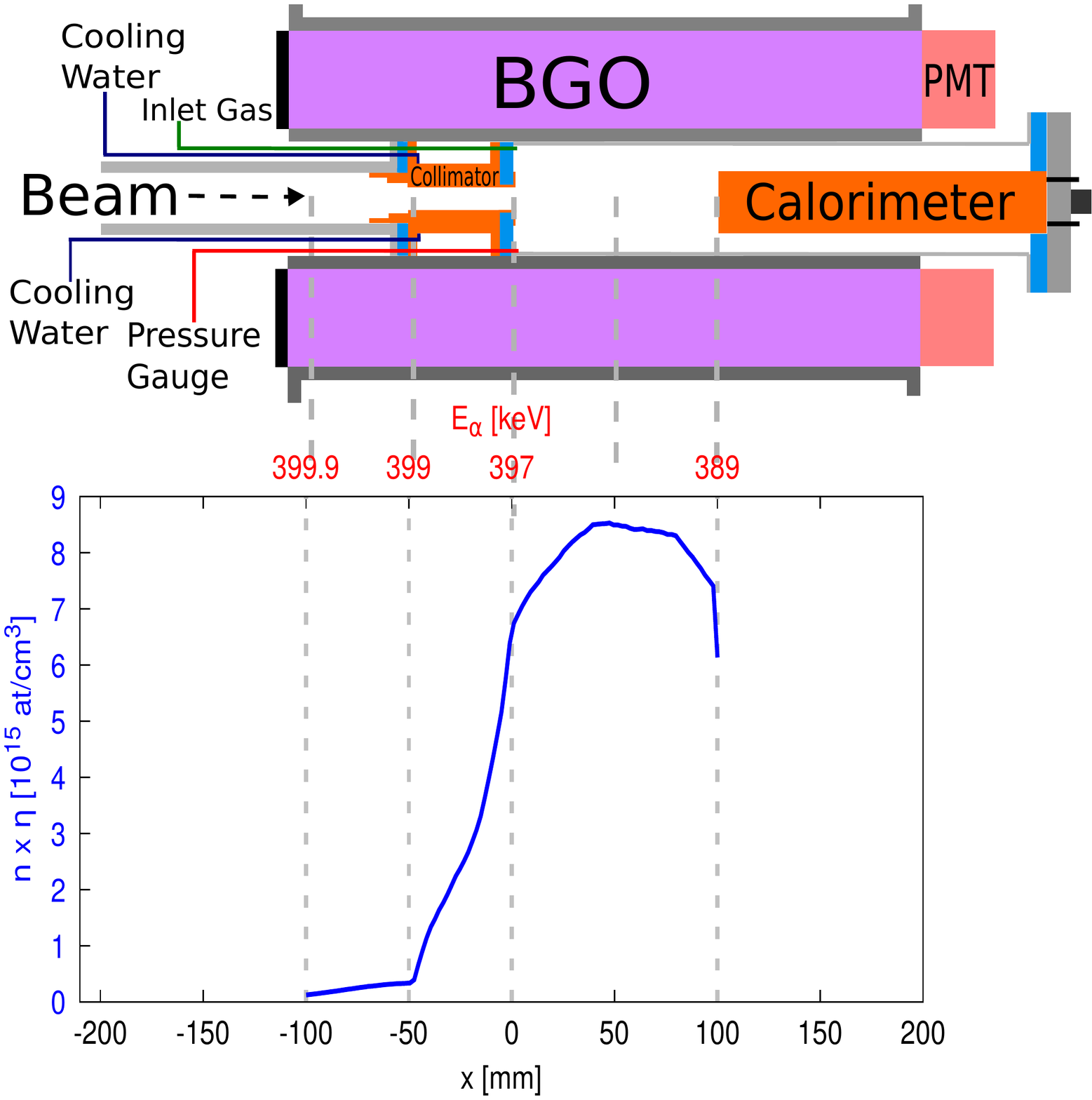}
    \caption{Schematic view of the experimental setup. Top panel:  Scattering chamber. Bottom panel: the product of target density and efficiency profiles, $n\,\times\,\eta$, with $x$ = 0 mm corresponding to the entrance of the chamber and $x$ = 50 mm the center. The top $x$-axis indicate the $^4$He$^+$ beam energy at selected positions $x$, indicated by the dark-grey dashed vertical lines. The highest efficiency is obtained for $E_\alpha$ = 397-389 keV.}
    \label{fig:setup}
\end{figure}

The experiment was performed at the Laboratory for Underground Nuclear Astrophysics (LUNA) 400 kV accelerator \cite{Broggini10-ARNPS,Broggini17-PPNP, Cavanna-2018IJMPA}, located in the deep underground INFN Laboratori Nazionali del Gran Sasso. Its location below the Gran Sasso massif reduces the natural background by three orders of magnitude in the 10-12 MeV $\gamma$-ray energy region of interest, enabling high-sensitivity studies \cite{Bemmerer05-EPJA,Boeltzig18-JPG}.

\subsection{Experimental setup and procedures}

The $^{4}$He$^{+}$ beam \cite{Formicola03-NIMA} of $E_{\alpha}$ = 399.9 keV (error smaller than 60 eV \cite{Formicola03-NIMA}) laboratory energy and 200-250 $\mu$A intensity was magnetically analyzed, collimated and drifted to a windowless gas target chamber \cite{Ferraro18-EPJAa} filled with 1.0\,mbar 99.9\% isotopically enriched  $^{22}$Ne gas of 99.995\% chemical purity (Figure \ref{fig:setup}). 

The chamber consisted of a  475 mm long stainless steel cylinder of 54 mm diameter. The central 100 mm of the cylinder formed the main target chamber. There, a constant pressure of 1.0 mbar was maintained by an MKS 248A solenoid valve controlled by an active feedback using a MKS Baratron 626 pressure gauge with 0.25\% precision \cite{Ferraro18-EPJAa}. Before reaching the main target chamber, the beam passed a 40 mm long, 7 mm diameter water-cooled collimator that ensured a pressure drop to the 10$^{-2}$ mbar range, obviating the need for a gas-tight entrance window. The pressure and temperature profiles have been precisely measured previously, so that the gas density inside the target chamber is known to 1.3\% precision \cite{Ferraro18-EPJAa}. Since the target chamber is windowless the precious enriched $^{22}$Ne gas was flushed from the collimator, collected and guided through a recirculation system by the vacuum pumping system. A noble-gas purifier (Monotorr PS4-MT3-R-2 with a PS4-C3-R-2 heated getter) removed possible nitrogen, oxygen and carbohydrates contaminations. The purified gas was re-used and entered the target chamber through a gas inlet. The gas purity over time was proved with the same setup in \cite{Ferraro18-EPJAa}

After passing the main target chamber, the beam was absorbed on a thick copper sheet that functions as the hot side of a beam calorimeter used for the beam intensity determination, with 1.5\% uncertainty \cite{Ferraro18-EPJAa}. 

The target chamber was surrounded by a 4$\pi$ bismuth germanate (BGO) borehole detector of 20$\,$cm outer and 6$\,$cm inner diameter and 28$\,$cm length \cite{Ferraro18-EPJAa}. This detector was optically divided into six segments, each of them read out by a dedicated photomultiplier tube (PMT). Each of the six signals was pre-amplified in an Ortec 113 preamplifier and digitized in a CAEN V1724 100 Ms/s 14 bit digitizer with a trapezoidal filter algorithm. Each channel was self-triggered. For the dead time determination, a pulser signal was connected to the test input of each preamplifier and also to a dedicated seventh acquisition chain, showing always less than 1\%  dead time. 

The $\gamma$-ray detection efficiency $\eta$ was calculated with a simulation of the present setup using the GEANT4 toolkit \cite{Agostinelli03-NIMA}, using the known $\gamma$-decay of the $E_{\rm x}$ = 10949 keV level \cite{Basunia-2016NDS}: 
\begin{itemize}
    \item $E$\textsubscript{$\gamma$}$\,=\,$6615.6$\,$keV, $E$\textsubscript{x}\textsuperscript{f}$\,=\,$4332.52$\,$keV, (10.81$\,\pm\,$1.03)\%
    \item $E$\textsubscript{$\gamma$}$\,=\,$7359.4$\,$keV, $E$\textsubscript{x}\textsuperscript{f}$\,=\,$3588.56$\,$keV, (4.69$\,\pm\,$0.47)\%
    \item $E$\textsubscript{$\gamma$}$\,=\,$8009.4$\,$keV, $E$\textsubscript{x}\textsuperscript{f}$\,=\,$2938.33$\,$keV, (13.56$\,\pm\,$1.32)\%
    \item $E$\textsubscript{$\gamma$}$\,=\,$9138.7$\,$keV, $E$\textsubscript{x}\textsuperscript{f}$\,=\,$1808.74$\,$keV, (57.21$\,\pm\,$3.43)\%
    \item $E$\textsubscript{$\gamma$}$\,=\,$10946.6$\,$keV, $E$\textsubscript{x}\textsuperscript{f}$\,=\,$0$\,$keV, (13.73$\,\pm\,$2.29)\%
    
\end{itemize} 
Further decays were taken from \cite{Basunia-2016NDS}.
The simulation was validated comparing experimental and simulated efficiencies measured with $\gamma$ calibration standard sources and the well known $E\,=\,278\,$keV resonance of the $^{14}$N(p,$\gamma$)$^{15}$O reaction \cite{Marta11-PRC}, the uncertainty is $\Delta \eta / \eta$ = 4\% \cite{Ferraro18-EPJAa}. The beam energy and the target pressure were selected in order to maximize the target density and the efficiency, according to the accelerator energetic range. At the beam energy corresponding to the resonance the reaction takes place effectively in a narrow interval around $x$ = 26$\,$mm, as obtained by energy loss calculation, slightly off the center of the target chamber. As a result, $\sim$40\% efficiency was obtained for $E_\alpha$ = 389-397 keV (= 329-336 keV in the center of mass system), corresponding to $E_{\rm x}$ = 10944-10951 keV excitation energy and, thus, covering most of the proposed resonance energies, see Table \ref{tab:soa}. For $E_\alpha$ = 397-400 keV, the efficiency was somewhat lower, 10-35\%, as expected due to the additional passive layers of the collimator and because of the increased distance between the $\gamma$-ray and detector interaction point with respect to the photomultiplier location.

\begin{table}
 \centering
 \caption{Experimental campaigns I and II: Running time $t$ in days, accumulated charge $Q$, $N_b$, namely the number of beam particles that impinged on target, and counts $N_{\rm ROI}$ in the region of interest for the on-resonance runs ($^4$He$^+$+$^{22}$Ne), for the no-beam background, and for the Ar+$^4$He$^+$ in-beam background run.}
    \resizebox{\columnwidth}{!}{%
   \begin{tabular}{l l r r c r}
\toprule
    \multicolumn{2}{l}{Campaign and run type} & $t$ [d] & $Q$ [C] & $N_b$ [$^{4}$He$^{+}$] & $N_{\rm ROI}$ \\
    \midrule
       I & $^{22}$Ne+$^4$He$^+$ & 16.2 & 312.5 & {2.0$\times10^{21}$}  & 26 \\
       I  & no-beam background  & 23.1 &   &  & 33 \\ 
      II & $^{22}$Ne+$^4$He$^+$ & 23.8 & 430.5 & 2.7$\times10^{21}$ & 8 \\
      II & Ar+$^4$He$^+$ & 7.2 & 75.6 & 4.7$\times10^{20}$ & 2 \\
      II & no-beam background &  40.1 &  &  & 18 \\
       \bottomrule
    \end{tabular}
    }%
    \label{tab:runs}
\end{table}

\subsection{Data taking campaign I}
\label{subsec:DataTakingCampaignI}

The data taking was performed in two subsequent campaigns, I and II (Table \ref{tab:runs}). In Campaign I, the BGO detector was used without external shielding.
During the experiment, for the  $^{22}$Ne+$^4$He$^+$ data taking, typically 12/24$\,$h long runs with set beam energy $E_{\alpha}$ = 399.9 keV were performed. Subsequently, the no-beam background was determined in dedicated runs.

The initial analysis of Campaign I gave a 0.4$\sigma$ excess in the net counting rate, 0.18$^{+0.40}_{-0.18}$ counts/day.
Given the result obtained in Campaign I, details are reported in Section ~\ref{sec:Analysis}, the shielding and background determination were improved for the subsequent Campaign II.

\subsection{Data taking campaign II}

\begin{figure*}[tb]
\centering
\includegraphics[width=\textwidth]{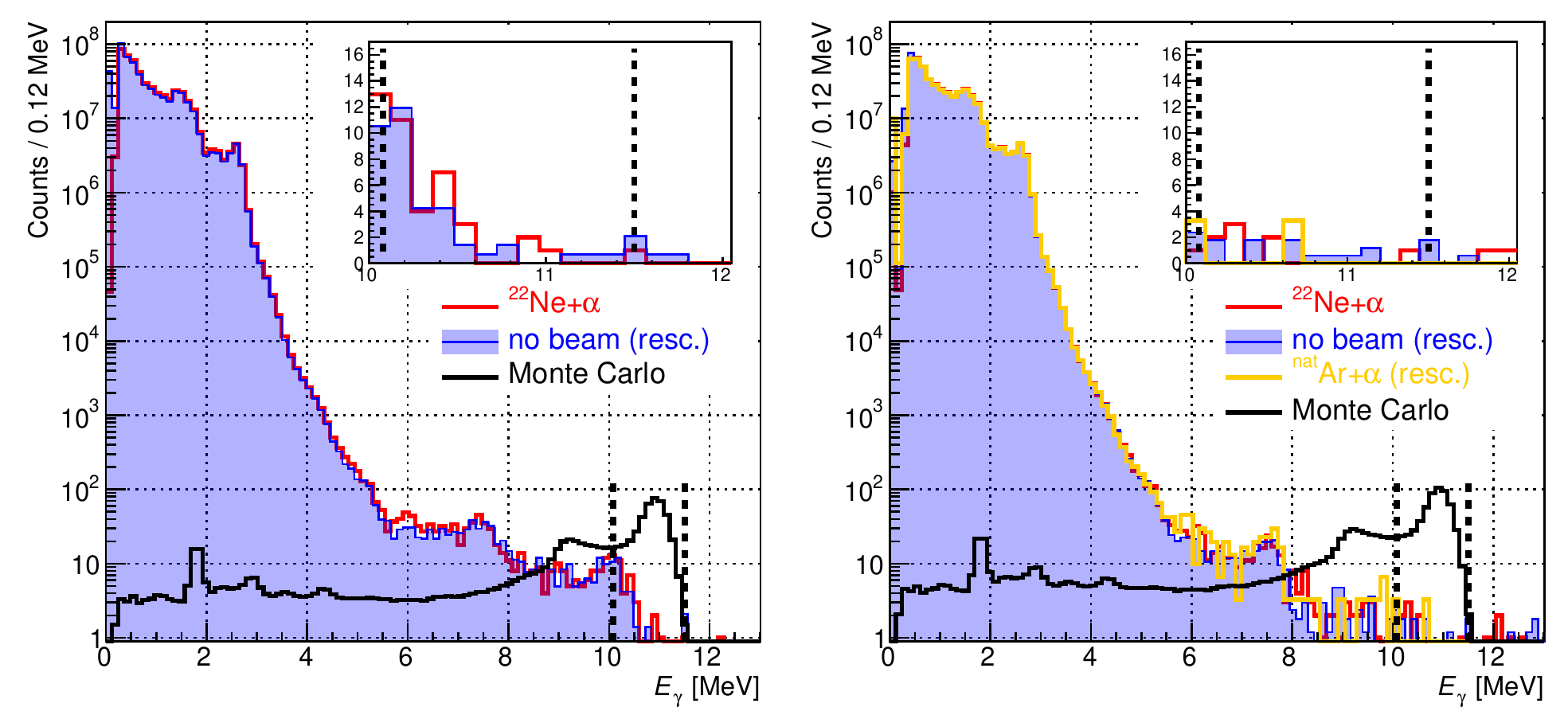}
    \caption{$\gamma$-ray spectra, add-back mode. Left, campaign I ($t$ = 16.2 d, $Q$ = 312.5 C); Right, campaign II ($t$ = 23.8 d, $Q$ = 430.5 C). Red line: $^4$He$^+$ + $^{22}$Ne run. Blue line: no-beam background (rescaled for equal time). The reduction of the background in 6-12 MeV region between campaign I and II is due to the shielding introduction. Orange line: $^4$He$^+$ + Ar beam induced background run (rescaled for equal time). Black line: Monte Carlo simulation for $\omega\gamma\,=\,3.6\times10^{-9}\,$eV, just to show the region of interest (ROI).
    The ROI is shown by vertical dashed lines. The inserts focuses on the ROI, in linear scale.
    }
    \label{fig:spectra}
\end{figure*}

In Campaign II, a 10 cm thick shield surrounded the BGO. It consisted of borated (5\%) polyethylene (PE-HMW 500 BOR5 by Profilan Kunstoffwerk) added outside the BGO. In the $E_\gamma$ = 6-18$\,$MeV region, the background in the BGO detector deep underground at LUNA was mainly given by neutron-induced effects that are believed to originate in ($\alpha,n$) neutrons caused by $\alpha$-decays in the natural decay chains in the rock \cite{Bemmerer05-EPJA, Best-2016NIMPA, Boeltzig18-JPG}. The shielding reduced the background counting rate in the region of interest by a factor of 3.4$\,\pm\,$0.3 (Table \ref{tab:runs}).
Note that Campaign II shows a lower signal in all parts of the spectrum despite the longer running times (Figure \ref{fig:spectra}).


In Campaign II, the $^{22}$Ne+$^4$He$^+$ runs were performed in a similar manner as in Campaign I. The no-beam background was determined for almost double the running time of the in-beam runs. In addition, for Campaign II also the in-beam background was determined by irradiating inert argon gas with the same $^4$He$^+$ beam as with neon gas. The argon was set to a pressure of 0.5 mbar, to reproduce, within 6\%, the same energy loss and angular beam straggling as in the $^{22}$Ne+$^4$He$^+$ case. Due to beam time constraints, only limited statistics was gained in the Ar+$^4$He$^+$ run.


\newcolumntype{d}[1]{D{.}{.}{#1}}

\section{Data analysis and results}
\label{sec:Analysis}

\subsection{Calibration and summing of $\gamma$-ray spectra}

The strong temperature dependence of the BGO scintillation efficiency causes resolution loss \cite{Melcher-1989NIMPB}, which was avoided in the present case by self-calibrating the time stamped list mode data, namely counts stored sequentially as time progresses, of each BGO segment (single mode) for each 12/24-hour run. The linear calibration was performed using the lines at 1461 and 2615 keV, that are due to the $^{40}$K and $^{208}$Tl decays in the room background, observable in each run.
An additional problem arose due to the large dynamic range covered by the signals, up to 12 MeV and beyond, where some non-linearity was observed in the photomultiplier gain. These effects were corrected based on spectra taken in the previous study of the $^{22}$Ne(p,$\gamma$)$^{23}$Na reaction \cite{Ferraro18-PRLb, Ferraro18-EPJAa}, that just preceded the present study. There, the same detector, chamber, and room temperature were used as in the current work. From these data \cite{Ferraro18-PRLb, Ferraro18-EPJAa}, known high-energy $\gamma$ lines from beam induced background reactions, namely the $^{11}$B(p,$\gamma$)$^{12}$C reaction ($E_{\gamma}\,=\,$4.4, 11.7, 16.1 MeV), the $^{13}$C(p,$\gamma$)$^{14}$N reaction ($E_{\gamma}\,=\,$7.9 MeV) and the $^{14}$N(p,$\gamma$)$^{15}$O reaction ($E_{\gamma}\,=\,$6.13 MeV) and from the $^{22}$Ne(p,$\gamma$)$^{23}$Na reaction (9.0 MeV) were available, bracketing the present region of interest in energy. The resultant non-linearity correction was 0.24$\,\pm\,$0.08$\,$MeV at $E_\gamma$ = 10.95$\,$MeV, corresponding to a 2.2\% effect. 

Next, for each run, the data from the six segments were combined offline. In each crystal, the pulser signal was gated out by requiring anticoincidence with a seventh channel only measuring the pulser. The time coincidence window was set at 3.5 $\mu$s \cite{Boeltzig18-JPG} both for gating out the pulser and to produce the add-back spectra corresponding to a virtual large detector  \cite{Boeltzig18-JPG,Caciolli11-AA}. In order to check whether the count rates of single runs of a given run type ($^4$He$^+$+$^{22}$Ne, $^4$He$^+$+Ar, or no beam) were mutually consistent and can, in fact, be added, a $\chi^2$ test \cite{Behnke13-Book} was performed for each campaign data set. For this test, the counting rate in the (n,$\gamma$)-dominated 6-20\,MeV energy region was used. There, a comparatively high counting rate of 39.6$\,\pm\,$1.3 counts/day (Campaign I) or 13.8$\,\pm\,$0.6 counts/day (Campaign II) ensured the applicability of the test. It showed a normal, non-skewed distribution for all five groups of runs studied, with $\chi^2/\nu$ values ranging from 0.7-1.1. The summed spectra for each run type and campaign are shown in Figure \ref{fig:spectra}.

\subsection{Interpretation of the experimental spectra}

The experimental spectra were compared with a Monte Carlo spectrum simulated using the highest upper limit for the resonance strength that is available in the literature, $\omega\gamma \le 3.6 \times 10^{-9}$\,eV \cite{Sallaska13-ApJSS,Iliadis10-NPA841_251}. The known decay scheme of the 10.95 MeV level goes mainly to the first excited state at 1.81 MeV \cite{Basunia-2016NDS}, but due to the summing effect of the 4$\pi$ BGO detector the dominant peak in the simulated add-back spectrum is at 10.95 MeV. For the analysis, a region of interest (ROI) of $E_\gamma \in [10.08; 11.50]$\,MeV was adopted (Figure \ref{fig:spectra}). This was obtained by both using simulation 
shown with a black line in Figure \ref{fig:spectra} 
and a devoted study on BGO resolution. The ROI also included the uncertainty due to the gain non-linearity. 

In the experimental $^{22}$Ne+$^4$He$^+$ spectra, no peak is observable in the region of interest (ROI). The net counts, obtained subtracting the background counts from the $^{22}$Ne+$^4$He$^+$ spectra, 
were always below the critical limits for
Campaign I and II respectively, i.e. the level
needed for a 95\% confidence level detection \cite{Gilmore08-Book}. 

These conclusions may, however, depend on the choice of the ROI. In order to study this possible effect, the analysis was repeated several times, shifting, in turn, both the lower and upper ROI limits in several steps by up to 550 keV, i.e. the full width at half maximum of a single $\gamma$-ray detected at these energies. For all these cases, the number of net counts in both campaigns remained below the 95\% confidence level critical limit for detection of a signal. For campaign I, the test analyses with ROI shifted to lower energies reproduced the non-significant, 0.4$\sigma$ excess described above for the final recommended ROI (\autoref{subsec:DataTakingCampaignI}) but the analyses with ROI shifted to higher energies showed no excess.

Therefore, to completely exclude any resonance detection, the shielding was introduced and the run time increased for Campaign II.

As an additional final check, both campaign single crystal spectra were checked when gating on add-back signals in the ROI \cite{Ferraro18-EPJAa}. The pattern expected from the Monte Carlo simulation for detection of the resonance in the single crystals, in particular the 9.14\,MeV $\gamma$ ray due to the decay to the first excited state of $^{26}$Mg, was not found. 

\subsection{Upper limit for the resonance strength under study}
\label{subsec:OmegaGammaUpperLimit}
The 334 keV resonance is isolated and narrow, meaning its total width $\Gamma\,=\,\Gamma_\alpha+\Gamma_\gamma$, which is in the present case of the order of 5 eV \cite{Longland09-PRC,Jayatissa-2020PhLB}, is much smaller than the difference $\Delta E$ to the nearest resonance, $\Gamma_\alpha+\Gamma_\gamma$ $\ll \Delta E$, and also than the target thickness when expressed in energy units of $\Delta E_{\rm targ} \sim 8$\,keV, thus $\Gamma_\alpha+\Gamma_\gamma$ $\ll \Delta E_{\rm targ}$. Therefore, in principle the resonance strength $\omega\gamma$ and the experimental yield $Y = N$\textsubscript{ROI}$ / \eta N_b$ as a function of background-subtracted counts $N$\textsubscript{ROI}, efficiency $\eta$, and impinging beam particles $N_b$, are connected by the so-called thick-target yield formula \cite{Iliadis15-Book}.

However, in the present case the energetically narrow, $\Delta E_{\rm beam} \sim 0.1$\,keV, beam from the accelerator significantly widens in energy due to beam energy straggling in the extended gas target \cite{Bemmerer18-EPL}, up to $\Delta E_{\rm strag} \sim 1.4$\,keV at the end of the target. The angular straggling has only a much smaller effect, $\leq$0.1 cm in lateral straggling over the entire target length. In addition the efficiency-corrected density profile $n\,\times\,\eta$ (Figure\ref{fig:setup}) deviates from an ideal box shape. The convolution of the beam energy distribution and the efficiency-corrected density profile produces a correction factor $C = (0.93\pm0.02)$ that modifies the ideal thick-target yield formula. The calculated correction takes into account the reduction of the yield due to the finite beam energy width and the target density profile \cite{Bemmerer18-EPL}.

The modified thick-target yield formula is then given by:
\begin{equation} \label{eq:ThickTargetYield}
\frac{N_{ROI}}{\eta N_b} =
Y = C\,\frac{\lambda_{\rm res}^2}{2} \frac{1}{\displaystyle\left.\frac{dE}{dx}\right|_{\rm eff}} \omega\gamma
\end{equation}
where $\lambda_{\rm res}^2/2$ = $3.64\times10^{-24}$\,cm$^2$ is the de Broglie wavelength at the resonance energy for $^{22}$Ne$+\alpha$ in the center-of-mass system and $\left.\frac{dE}{dx}\right|_{\rm eff}$ = $32.5\times10^{-15}$ eV cm$^2$ is the effective stopping power at the resonance energy, as given by the SRIM software package \cite{Ziegler10-NIMB}, again in the center-of-mass system. The sought-after resonance strength $\omega\gamma$ is then obtained by solving Equation (\ref{eq:ThickTargetYield}).
\begin{figure}[tb]
\centering
\includegraphics[width=\columnwidth]{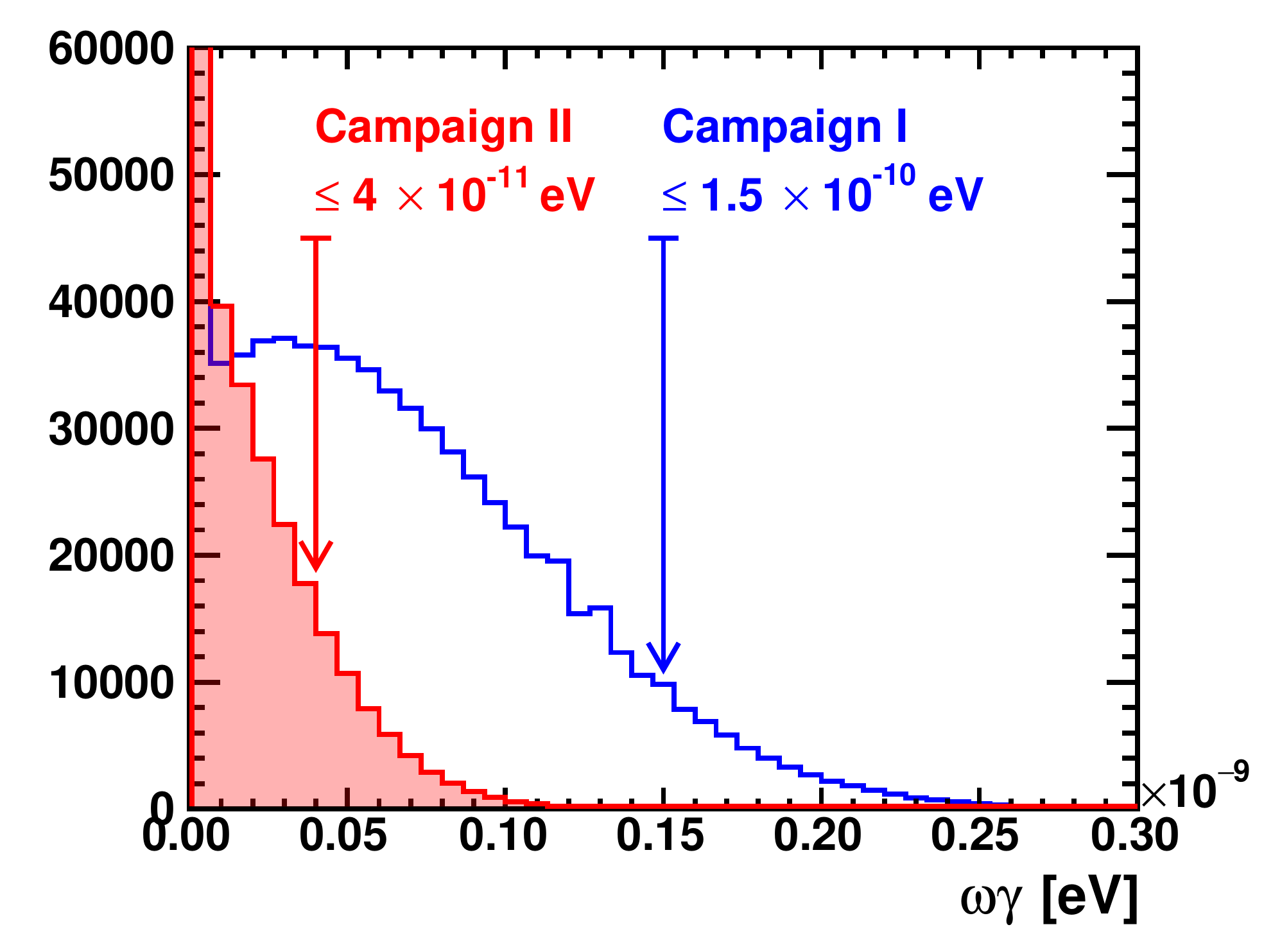}
\caption{Probability density functions of the resonance strength $\omega\gamma$ for campaign I (blue) and campaign II (red).  Upper limits at 90\% confidence level are denoted by arrows.}
    \label{fig:pdf}
\end{figure}

To obtain an upper limit for the resonance strength $\omega\gamma$, a Monte Carlo sampling technique modeled on that described by \cite{Rolke01-NIMA} was used for both campaign data. The background and the signal, observed in the same ROI but in background and $^{22}$Ne$\,+\,\alpha$ spectra respectively, were assumed to follow a Poisson distribution, and for each of the two campaigns separately, $10^6$ samples were taken from the $^{22}$Ne+$^4$He$^+$ and no-beam count rates. 

For Campaign I, the resultant probability density function for the net counts actually shows a maximum slightly in excess of zero, but only at 0.4$\sigma$ confidence level (Figure \ref{fig:pdf}). Campaign II shows no maximum above zero. For the Figure, the counts were converted to $\omega\gamma$ values by eq.\ref{eq:ThickTargetYield}. Some samples (38\% and 79\% of samples for campaigns I and II, respectively) resulted in unphysical negative net counts; they were forced to zero and included in the plot and upper limit determination. 

From the probability density functions, considering the already mentioned $C$ correction factor (Equation \ref{eq:ThickTargetYield}), upper limits at 90\% confidence level of $\le 1.5 \times 10^{-10}$ eV and $\le 4.0 \times 10^{-11}$ eV are obtained for Campaigns I and II, respectively. 

\section{Estimated sensitivity for higher-energy resonances}
\label{sec:Future}

The present experiment used the highest beam energy available at the present LUNA 400 kV accelerator, meaning no higher-energy $^{22}$Ne($\alpha,\gamma$)$^{26}$Mg resonances could be studied. Experiments at higher energies can instead be conducted at the new LUNA-MV 3.5 MV accelerator deep underground in the Gran Sasso laboratory \cite{Ferraro21-FASS, Broggini19-NCA, Sen-2019NIMPB} or at other underground accelerators. 
Therefore, the sensitivity of the present setup and approach for the study of higher-energy resonances in $^{22}$Ne($\alpha,\gamma$)$^{26}$Mg is discussed in the following.

\begin{figure}[tb]
\includegraphics[width=\columnwidth]{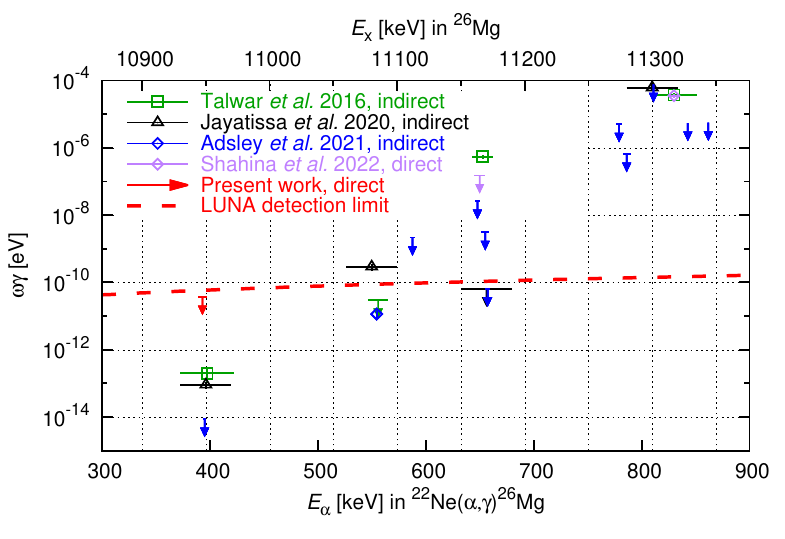}
\caption{\label{fig:Piatti_Sensitivity} Sensitivity of the present setup for the direct detection of $^{22}$Ne($\alpha,\gamma$)$^{26}$Mg resonances, as a function of laboratory resonance energy $E_\alpha$, see text for details. Previous indirect data and upper limits (indicated by arrows) from \cite{Talwar16-PRC} (in green), \cite{Jayatissa-2020PhLB} (in black) and \cite{Adsley-21} (in blue) are shown, as well as the present experimental upper limit,in red. A very recent new data point and limit are shown in purple \cite{2022PhRvC.106b5805S}, see Note added in Proof.}
\end{figure}

Using the measured background in Campaign II (Table \ref{tab:runs}), an average $^4$He$^+$ beam current of 300\,$\mu$A \cite{Sen-2019NIMPB}, and a running time of 40 days per data point, the sensitivity of the present setup to positively detect a resonance with 90\% confidence level \cite{Gilmore08-Book} has been calculated for a number of higher-energy resonances ( Fig.\ref{fig:Piatti_Sensitivity}).

In relation to the other two possible resonances discussed in the introduction: For the $E$ = 469 keV ($E_{\rm x}$ = 11084 keV) resonance, the indirect positive value of $(2.8 \pm 0.8) \times 10^{-10}$ eV \cite{Jayatissa-2020PhLB} would be detectable in the present setup, while for the indirect upper limit of $\le 2.95 \times 10^{-11}$ eV  \cite{Talwar16-PRC}, a direct upper limit of similar size would be possible. For the $E$ = 556 keV ($E_{\rm x}$ = 11171 keV) resonance, the indirect value of $(5.4 \pm 0.7) \times 10^{-7}$ eV \cite{Talwar16-PRC} would be detectable, and the indirect upper limit of $\le 6 \times 10^{-11}$ eV  \cite{Jayatissa-2020PhLB} again within reach of a direct experiment (Figure \ref{fig:Piatti_Sensitivity}).
For resonances that are well above the $E_\alpha$ = 565\,keV threshold for the $^{22}$Ne($\alpha,n$)$^{25}$Mg reaction, the present sensitivities are only applicable if the neutron partial width $\Gamma_n$ of the corresponding level \cite{Massimi17-PhLB} is smaller than the $\gamma$ width, $\Gamma_n\,<\,\Gamma_{\gamma}$.
The present sensitivities (Figure \ref{fig:Piatti_Sensitivity}) may also be relevant to other underground facilities, depending on the background situation \cite{Olivas-Gomez-2012A&A,
Zhang-PhysRevLett2021,
Bemmerer18-SNC}.

\section{Summary and conclusions}
\label{sec:Summary}
The $^{22}$Ne($\alpha,\gamma$)$^{26}$Mg resonance at $E$ = 334 keV in the center-of-mass system ($E_x$ = 10949 keV excitation energy in $^{26}$Mg) has been studied in a direct experiment. Using the LUNA 400 kV accelerator deep underground in the Gran Sasso laboratory, a windowless, isotopically enriched $^{22}$Ne gas target and a 4$\pi$ BGO summing detector, an experimental upper limit of 4.0$\times10^{-11}$ eV (90\% confidence level) has been derived for the strength of this resonance. 
The new limit is higher than most of the previous limits obtained by indirect methods \cite{Longland12-PRC,Talwar16-PRC,Massimi17-PhLB,Ota20-PLB,Adsley-21}, but it is the first direct result. 
The present approach may be extended to higher energies at the upcoming LUNA-MV 3.5 MV accelerator deep underground in the Gran Sasso laboratory \cite{Ferraro21-FASS, Broggini19-NCA, Sen-2019NIMPB}.\\
\newline{}

\subsection*{Acknowledgements} D. Ciccotti and the technical staff of the LNGS are
gratefully acknowledged for their help.
Financial support by INFN, the
Italian Ministry of Education, University and Research (MIUR) through the "Dipartimenti di eccellenza" project "Science of the Universe", the European Union (ERC Consolidator Grant project {\em STARKEY}, no. 615604, ERC-StG SHADES, no. 852016), and ChETEC-INFRA, no. 101008324), Deutsche For\-schungs\-gemeinschaft (DFG, BE 4100-4/1), the
Helm\-holtz Association (ERC-RA-
0016), the Hungarian National Research, Development and
Innovation Office (NKFIH K134197), the European Collaboration for Science and Technology (COST Action ChETEC, CA16117) and a
DAAD fellowship at HZDR for D.P. are
gratefully acknowledged. C. G. B., T. C., T. D. and M. A. acknowledge funding by STFC UK (grant no. ST/L005824/1).

For the purpose of open access, the authors have applied a Creative Commons Attribution (CC BY) licence to any Author Accepted Manuscript version arising from this submission.

\subsection*{Note added in proof}

After the final version of the present manuscript was completed, the authors became aware of a very recent, new direct experiment \cite{2022PhRvC.106b5805S} showing a new value for the $E_\alpha$ = 830 keV resonance strength and a new upper limit for the 653 keV resonance. These new literature data have been added to Figure \ref{fig:Piatti_Sensitivity}.

  

\end{document}